\documentclass{spie}  

\def\titlefont{\LARGE\bf}  

\usepackage{fancyhdr} 
\usepackage{time}

\usepackage{url}
\usepackage{graphicx}
\usepackage{rotating,rotfloat}
\usepackage{wrapfig} 
\usepackage{subfigure}
\usepackage{amsmath}
\usepackage{amssymb}

\usepackage{txfonts}

\usepackage{epic}
\usepackage{eepic}

\def\abs#1{|#1|}

\def\secref#1{Sect.\@~\ref{#1}}

\def\eqref#1{Eq.\@~(\ref{#1})}
\def\eqsref#1{Eqs.\@~(\ref{#1})}

\def\figref#1{Fig.\@~\ref{#1}}

\def\I{{\bf I}} 
\def\A{{\bf A}} 
\def\C{{\bf C}} 
 
\def\K{{\bf Q}} 
\def\Q{{\bf Q}} 
\def\R{{\bf R}}

\def\a{\mbox{\boldmath$\alpha$}} 
\def\b{{\bf b}} 
\def\B{\mbox{\boldmath$\beta$}} 
\def\d{{\bf d}} 
\def\da{{\delta\a}}  

\def\ip#1#2{\left\langle#1\,,\,#2\right\rangle}

\def\Re{\operatorname{Re}}
\def\Im{\operatorname{Im}}

\def\Fourier{\operatorname{\mathfrak{F}}}

\def\bar#1{\overline{#1}}

\title{Multi-frame blind deconvolution with linear equality constraints}

\author{Mats G. L{\"o}fdahl\skiplinehalf
  Institute for Solar Physics of the Royal Swedish Academy of
  Sciences}

\authorinfo{Further author information: E-mail:
  \url{mats@astro.su.se}, WWW: \url{www.solarphysics.kva.se/~mats}, Postal
  address: Institute for Solar Physics, AlbaNova University Center, SE--106\,91 Stockholm,
  Sweden}

\fancypagestyle{firstpage}{\chead{\vbox to 0mm{\vss\scriptsize To
      appear in Proc.\ SPIE \textbf{4792-21},
      \textit{Image Reconstruction from Incomplete
        Data II},  Bones, Fiddy \& Millane, eds., Seattle,
      Washington, USA, July 2002.
  \vspace{5mm}}}
  }

\begin{document} 

\maketitle 
\thispagestyle{firstpage} 

\begin{abstract} 
  
  The Phase Diverse Speckle (PDS) problem is formulated mathematically
  as Multi Frame Blind Deconvolution (MFBD) together with a set of
  Linear Equality Constraints (LECs) on the wavefront expansion
  parameters.  This MFBD--LEC formulation is quite general and, in
  addition to PDS, it allows the same code to handle a variety of
  different data collection schemes specified as data, the LECs,
  rather than in the code. It also relieves us from having to derive
  new expressions for the gradient of the wavefront parameter vector
  for each type of data set. The idea is first presented with a simple
  formulation that accommodates Phase Diversity, Phase Diverse
  Speckle, and Shack--Hartmann wavefront sensing. Then various
  generalizations are discussed, that allows many other types of data
  sets to be handled.
  
  Background: Unless auxiliary information is used, the Blind
  Deconvolution problem for a single frame is not well posed because
  the object and PSF information in a data frame cannot be separated.
  There are different ways of bringing auxiliary information to bear
  on the problem. MFBD uses several frames which helps somewhat,
  because the solutions are constrained by a requirement that the
  object be the same, but is often not enough to get useful results
  without further constraints.  One class of MFBD methods constrain
  the solutions by requiring that the PSFs correspond to wavefronts
  over a certain pupil geometry, expanded in a finite basis. This is
  an effective approach but there is still a problem of uniqueness in
  that different phases can give the same PSF.  Phase Diversity and
  the more general PDS methods are special cases of this class of
  MFBD, where the observations are usually arranged so that in-focus
  data is collected together with intentionally defocused data, where
  information on the object is sacrificed for more information on the
  aberrations. The known differences and similarities between the
  phases are used to get better estimates.
  
\end{abstract}


\section{Introduction}
\label{sec:intro}  

We first present a technique for jointly estimating the common object
and the aberrations in a series of images that differ only in the
aberrations. With no extra information beyond the image formation
model, including aberrations from the phase in the generalized pupil
transmission function, it is a Maximum-Likelihood (ML) Multi-Frame
Blind Deconvolution (MFBD) \cite{schulz93multi-frame} method.
Constraining the PSFs to be physical by requiring that they come from
an underlying parameterization of the phase over the pupil is a
powerful technique. Although such methods do work
\cite{schulz97multiframe,kampen98multi-frame}, methods using more
information work better\cite{tyler98comparison}. Data sets used in
Phase-Diverse Speckle (PDS) interferometry
\cite{paxman92phase,lofdahl94wavefront,paxman96evaluation} has such
extra information in the form of two (or more) imaging channels with a
known difference in phase (at least to type), with Phase Diversity
(PD) \cite{gonsalves82phase,paxman92joint,lofdahl94wavefront} as the
special case of only one such pair (or set). These are all different
data collection schemes lending themselves to similar joint inversion
techniques. The purpose of the formulation presented here is to
recognize the similarities and outlining a method for treating them
all with a single algorithm.  We discuss how the formulation
accommodates also Shack--Hartmann (SH) wavefront sensing (WFS) as well
as, with a couple of simple generalizations, several other types of
data sets.

We start with the ML metric for a MFBD data set and derive an
algorithm for the joint estimation of object(s) and aberrations for a
variety of imaging scenarios. In its simplest form, it is equivalent
to MFBD, while the known relations between image channels for
particular data collection schemes are expressed as Linear Equality
Constraints (LECs).

Throughout this paper a Gaussian noise model for the data is assumed
because that permits simplifications in the ML solution methods. It is
our experience with Gaussian noise PD, that it is a good model for
low-contrast objects like the solar photosphere
\cite{lofdahl94wavefront} but also that artificial high-contrast
objects (like a laser--pinhole source) can often be imaged in such a
way that the noise is negligible\cite{lofdahl00calibration}.

A code based on the formulation presented here has been successfully
applied to PD, PDS, and MFBD data of various kind, both from the
Swedish Vacuum Solar Telescope as well as several different simulated
data sets.

\section{Multi-Frame Blind Deconvolution}
\label{sec:notation}

\subsection{Forward Model and Error Metric}
\label{sec:model}

We use an isoplanatic image formation model with Gaussian white noise,
which means that we assume that the optical system can be characterized
by a generalized pupil function, which can be written, for an image
frame with number $j\in\{1,\ldots,J\}$, as
\begin{equation}
  \label{eq:pupil}
  P_j = A_j\exp\{i\phi_j\}
\end{equation}
where $\phi_j$ is the phase and $A_j$ is a binary function that
specifies the geometrical extent of the corresponding pupil.  A data
frame $d_j$ can then be expressed as the convolution of an object,
$f$, and a point spread function (PSF), $s_j =
\abs{\Fourier^{-1}\{P_j\}}^2$.  In the Fourier domain we get
\begin{equation}
  \label{eq:image-formation}
  D_j(u) = F(u) \cdot S_j(u) + N_j(u),
\end{equation}
where $S_j$ is the OTF, $N_j$ is an additive noise term with Gaussian
statistics and $u$ is the 2-D spatial frequency coordinate. For
brevity, we will drop this coordinate for the remainder of the paper.

We parameterize the unknown phases by expanding them in a suitable
basis, $\{\psi_m\}$, allowing for a part of the phase, $\theta_j$, to
be excepted from the expansion,
\begin{equation}
  \label{eq:phi}
  \phi_{j} = \theta_{j} + \sum_m^M \alpha_{jm} \psi_{m}; \qquad
  \forall j.
\end{equation}

The Gaussian white noise assumption allows us to use the inverse
Wiener filter estimate of the object,
\begin{equation}
  \label{eq:F}
  F = \frac{1}{Q}\sum_j S^*_jD_j,
\end{equation}
where 
\begin{equation}
  \label{eq:Q}
  Q = \gamma_{\rm obj} + \sum_{j}^J\abs{S_{j}}^2,
\end{equation}
to derive a metric in a form that does not explicitly involve the
object\cite{gonsalves82phase} and that has been shown to correspond to
a ML estimate of the phases\cite{paxman92joint}. With two
regularization parameters\cite{vogel98fast}, this metric can be written
as
\begin{equation}
  \label{eq:L-mfbd}
  L(\a) = \sum_u\biggl[
  \sum_j^J 
  \abs{D_{j}}^2 -
  \frac{\abs{\sum_j^J D^*_{j} S_{j}}^2}{Q}
  \biggr]
  + \frac{\gamma_{\rm wf}}{2}\sum_m^M\frac{1}{\lambda_m}\sum_j^J \abs{\alpha_{jm}}^2.
\end{equation}
When minimizing $L$, the $\gamma_{\rm obj}$ term in $Q$ has the
effect of establishing stability with respect to perturbations in the
object. Its use in \eqref{eq:F} suggests setting $\gamma_{\rm obj}$
to something proportional to the inverse of the signal to noise ratios
of the image data frames.  The other regularization parameter,
$\gamma_{\rm wf}$, stabilizes the wavefront estimates and can be set
by examining the relation between $L$ and the RMS of the
wavefront\cite{engl96regularization}. For simplicity, this term is
given under the assumption that the $\psi_m$ are Karhunen--Lo\`eve
(KL)\cite{roddier90atmospheric} functions, where $\lambda_m$ is the
expected variance of mode $m$. This should work reasonably well also
for low-order Zernike polynomials. In the general case, the wavefront
regularization term is a matrix operation involving the
covariances.\cite{vogel98fast}

Note that, although presented in a PD setting, this metric has nothing
to do with PD per se, it is just a MFBD metric. The PD part enters in
the parameterization in earlier PD methods and, equivalently, in the
LECs in this presentation.

\subsection{Gradient and Hessian for Traditional Phase Diversity}
\label{sec:trad-phase-diversity}

Efficient minimization of $L$ requires the gradients and, for some
methods, the Hessian with respect to the aberration parameters.
We start by considering the traditional PD problem, for which the
normal equations can be written as
\begin{equation}
  \label{eq:L-minimization-constrained-PD}
  \A^{\rm PD}\cdot\da - \b^{\rm PD} \simeq 0,
\end{equation} 
where the elements of $\a$ are the coefficients of an expansion of the
single phase that is at the pupil of all diversity channels.
The $M$ elements of $\b^{\rm PD}$ can be expressed as the Euclidean
inner products, $\ip{\cdot}{\cdot}$, of an expression for the gradient
of $L$ and the aberration basis functions taken over the definition
area of the basis functions,\cite{vogel98fast}
\begin{equation}
  \label{eq:rhs}
  b^{\rm PD}_{m} = \ip{-2 \sum_{j=1}^J \Im\Bigl[P_j^* \Fourier\Bigl\{ p_j
  \Re\bigl[\Fourier^{-1}\{F^* D_j - \abs{F}^2 S_j\}\bigr] \Bigr\}\Bigr]}{\psi_{m}}  
  + \gamma_{\rm wf}\frac{\alpha_{m}}{\lambda_m}.
\end{equation}
An approximation of the $M \times M$ Hessian matrix $\A^{\rm PD}$ can
be derived if $Q$ is regarded as a fixed quantity\cite{vogel98fast}.
The elements can then be written as
\begin{equation}
  \label{eq:hess-PD}
  A^{\rm PD}_{m'm} = 
  \ip{4\sum_{j=1}^{J-1} \sum_{j'=j+1}^{J} 
  \Im\biggl[ 
  P^*_{j}\Fourier\biggl\{
  p_{j}\Fourier^{-1}\Bigl\{\frac{D^*_{j'}}{Q} U_{jj'} \Bigr\}
  \biggr\} -
  P^*_{j'}\Fourier\biggl\{
  p_{j'}\Fourier^{-1}\Bigl\{\frac{D^*_{j}}{Q} U_{j'j} \Bigr\}
  \biggr\} 
  \biggr]}{\psi_{m}},
\end{equation}
where
\begin{equation}
  \label{eq:Uij}
  U_{ij} =
   \frac{D_{i}}{Q} \Fourier\Bigl\{\Im\bigl[p^*_{j}\Fourier^{-1}\{\psi_{m'}P_{j}\}\bigr]\Bigr\}
  - \frac{D_{j}}{Q}\Fourier\Bigl\{\Im\bigl[p^*_{i}\Fourier^{-1}\{\psi_{m'}P_{i}\}\bigr]\Bigr\} .
\end{equation}

Other PD gradient and Hessian formulae are also available in the
literature\cite{paxman92joint,lofdahl94wavefront}. We base our
derivations on the ones given here, because they incorporate the
regularization terms.

\subsection{Gradient and Hessian for MFBD}
\label{sec:gradient}

The difference between MFBD and PD is that we in general do not know
any relation between the wavefronts in the different channels. The PD
gradient and Hessian expressions can then be simplified considerably
by relaxing all dependencies between imaging channels, because we can
then set all multiplications involving pupil quantities with different
$j$ to zero.  This corresponds to $J$ different pupils with
independent phases.

We also have to multiply with a factor $\sqrt{J}$ for each
differentiating operation. This is because we are splitting each
independent variable $\alpha_m$ into $J$ different $\alpha_{jm}$. The
vector sum of $J$ instances of identical d$\alpha_{jm}$ is $\sqrt{J}$
times larger than d$\alpha_m$.  Because this factor appears in the
denominator of the derivatives, we have to correct the gradient and
the Hessian by multiplying with $\sqrt{J}$ and $J$, respectively.

We lexicographically arrange all the $\alpha_{jm}$ in a
single column vector, $\a$, with $N=JM$ elements, so that we can also
refer to them as $\alpha_n$, with a single index $n=(j-1)M + m$,
\begin{equation}
  \label{eq:array-defs}
  \a = \begin{bmatrix}
    \alpha_{11} & \alpha_{12} &\dots & \alpha_{JM}
  \end{bmatrix}^{\rm T} 
        =  \begin{bmatrix}
    \alpha_{1} & \dots&  \alpha_{N}
  \end{bmatrix}^{\rm T}.
\end{equation}
We can then write the normal equations of the MFBD problem as
\begin{equation}
  \label{eq:L-minimization-constrained}
  \A^{\rm MFBD}\cdot\da - \b^{\rm MFBD} \simeq 0,
\end{equation}
where the inner product part of \eqref{eq:rhs} simplifies to the
individual terms of the summations,
\begin{equation}
  \label{eq:rhs2}
  b^{\rm MFBD}_{jm} = \ip{-2\sqrt{J}\Im\Bigl[P_j^* 
    \Fourier\bigl\{ p_j \Re[\Fourier^{-1}\{F^* D_j - \abs{F}^2 S_j\}]\bigr\}\Bigr]}{\psi_{m}}  
  + \gamma_{\rm wf}\frac{\alpha_{jm}}{\lambda_m}.
\end{equation}
Due to the independence between channels, the $N\times N$ Hessian
$\A^{\rm MFBD}$ is block-diagonal, 
\begin{equation}
  \label{eq:Amfbd+bmfbd}
  \A^{\rm MFBD} 
         = \begin{bmatrix}\A_1\\&\ddots\\&&\A_J\end{bmatrix},
\end{equation}
where each $M\times M$ matrix $\A_j$ has $A^{\rm MFBD}_{jm'm}$ as its
element at $(m',m)$.  With some arithmetics, \eqref{eq:hess-PD} can be
simplified so that these elements can be written as
\begin{equation}
  \label{eq:H_psi2}
  \begin{split}
    A^{\rm MFBD}_{jm'm} &=  4J\left\langle\Im\biggl[\,\, \sum_{j'=1}^{j-1}    
      \biggl(
      P^*_{j'}\Fourier\biggl\{p_{j'} \Fourier^{-1}\biggl\{\frac{
        D^*_{j}}{Q} V_{j'j} \biggr\} \biggr\} 
      +
      P^*_{j}\Fourier\biggl\{p_{j} \Fourier^{-1}\biggl\{\frac{
        D^*_{j'}}{Q} V_{j'j}\biggr\} \biggr\}
      \biggr)
      +\mbox{}\right.
    \\
    &\mbox{}\hspace{55pt} - \left.  \sum_{j'=j+1}^{J}   
      \biggl(
      P^*_{j'}\Fourier\biggl\{p_{j'} \Fourier^{-1}\biggl\{\frac{
        D^*_{j}}{Q} V_{j'j} \biggr\} \biggr\} 
      +
      P^*_{j}\Fourier\biggl\{p_{j} \Fourier^{-1}\biggl\{\frac{
        D^*_{j'}}{Q} V_{j'j}\biggr\} \biggr\}  
      \biggr)
      \biggr] \,,\, \psi_{m}\right\rangle,
  \end{split}
\end{equation}
where
\begin{equation}
  \label{eq:Xj}
  V_{ij} =  \frac{D_{i}}{Q}
  \Fourier\bigl\{\Im\bigl[p^*_{j}\Fourier^{-1}\{\psi_{m'} P_{j}\}\bigr]\bigr\}.
\end{equation}

\section{Linear Equality Constraints}
\label{sec:LEC}

\subsection{Notation and Theory}
\label{sec:theory}

Now we add information about the data set in the form of LECs. The
theory for solving optimization problems with LECs can be found in
some text books on numerical methods, e.g.
Ref.\@~\citenum{kahaner89numerical}.  The idea is that each constraint
reduces the number of unknowns by one. We can transform a constrained
optimization problem with $N$ parameters and $N_C$ constraints into an
unconstrained problem with $N-N_C$ unknowns.

The constraints are given as a set of linear equations that have to be
satisfied exactly,
\begin{equation}
  \label{eq:constraint-equation}
  \C\cdot \a - \d  = 0,
\end{equation}
while minimizing $L$. With $N_{\rm C}$ linearly independent
constraints, $\C$ is an $N_{\rm C}\times N$ matrix, where $N_{\rm C}
< N$. This means \eqref{eq:constraint-equation} is under-determined
and can therefore be solved exactly by several different $\a$.  All
solutions to \eqref{eq:constraint-equation} can be written as
\begin{equation}
  \label{eq:subst}
  \a = \bar\a + \K_2\cdot\B,
\end{equation}
where $\bar\a$ is a particular solution to
\eqref{eq:constraint-equation} and the $N'=N-N_{\rm C}$ column
vectors of $\K_2$ are an orthogonal basis of the null space of $\C$.
A particular solution can always be found by setting the $N'$ last
elements of $\bar\a$ to zero and solving the reduced system
\begin{equation}
  \label{eq:bar-a}
  \C'\bar\a' - \d = 0,
\end{equation}
where $\C'$ is the upper left $N_C\times N_C$ submatrix of $\C$ and
$\bar\a'$ is a vector containing the first $N_C$ elements of $\bar\a$.
Note, though, that all known differences can be incorporated into
$\theta_j$, so we can set $\d\equiv0$ without loss of generality, and
we always have the particular solution $\bar\a=0$.
We can find a basis for the null space by using an orthogonal
decomposition of $\C$, such as SVD or QR.  With QR factorization, we
write
\begin{equation}
  \label{eq:QR}
  \C^{\rm T}=\Q\cdot\R
  ;\qquad
  \Q=\begin{bmatrix}
    \K_1 & \K_2
  \end{bmatrix},
\end{equation}
where $\K_2$ is the rightmost $N'$ columns of $\K$ and T used as a
superscript denotes matrix transpose.

The constrained minimization problem in $\a$ can be transformed into
an unconstrained minimization problem in a reduced set of variables,
$\B$, an $N'$ element vector of parameters, $\beta_{n'}$.  The normal
equations for the transformed problem are obtained by left-multiplying
\eqref{eq:L-minimization-constrained} with $\K_2^{\rm T}$ and
substituting $\K_2\delta\B$ (\eqref{eq:subst}, $\bar\a=0$) for
$\delta\a$,
\begin{equation}
  \label{eq:L-minimization-unconstrained}
  \K_2^{\rm T}\A^{\rm MFBD}\K_2\cdot\delta\B - \K_2^{\rm
  T}\cdot\b^{\rm MFBD} \simeq 0.
\end{equation}
Once we have a solution for $\B$, we easily get the solution for $\a$
from \eqref{eq:subst}.

In order to efficiently express the LECs, we need to be able to
distinguish between frames collected in different ways. We will
therefore expand the $j$ index to a set of two indices, $k$ and $t$,
so that $j = k + (t-1)K \in\{1,\ldots,TK\}$. We will use
$k\in\{1,\ldots,K\}$ for simultaneous exposures in diversity channels
and $t\in\{1,\ldots,T\}$ for discrete time or different realizations
of atmospheric turbulence. With $K=1$ we have MFBD, while $T=1$
corresponds to PD. If $K=T=1$, we have the BD problem.


\subsection{Phase Diversity with only Known Differences in Phase}
\label{sec:PD-known}

\begin{figure}[htbp]
  \centering
  \hfill
  \begin{minipage}[b]{.47\textwidth}
    \centering
    \begin{minipage}[t]{.7\textwidth}
      \centering
      \begin{minipage}[t]{.6\textwidth}
        \vss
        \includegraphics[angle=-90,bb=0 35 505 710,width=\textwidth]
        {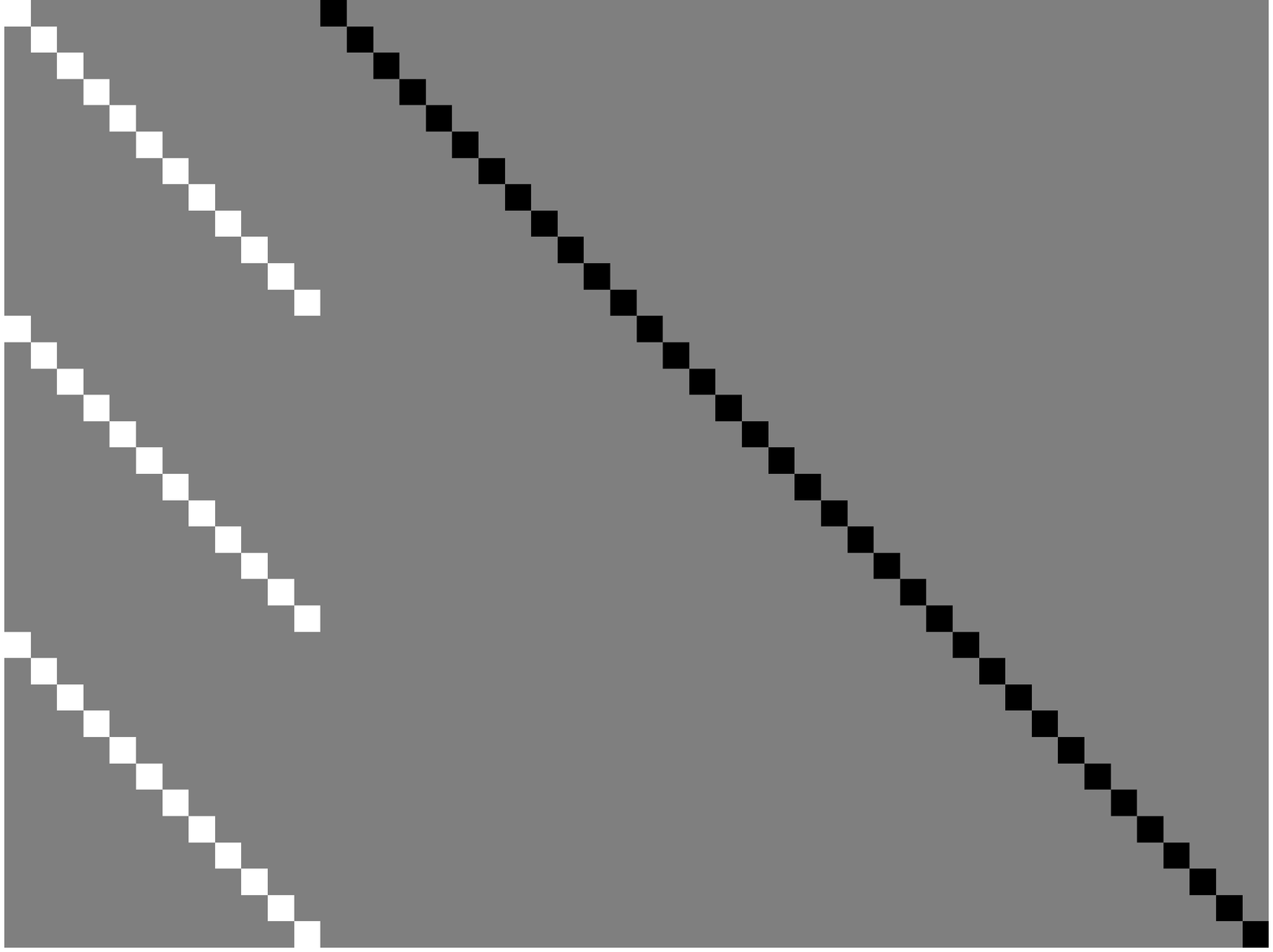}
        \vskip.15\abovecaptionskip
        \hbox{\footnotesize\textbf{(a)} $\C^{\rm PD'}$}
        \vskip\belowcaptionskip
        \vfill
        \vskip-.2\abovecaptionskip
        \hfill\hbox{\footnotesize\textbf{(b)} ${\K}_2$\ }
        \vskip\belowcaptionskip
      \end{minipage}
      \
      \begin{minipage}[t]{.15\textwidth}
        \vss
        \includegraphics[angle=-90,bb=0 35 505 163,width=\textwidth]
        {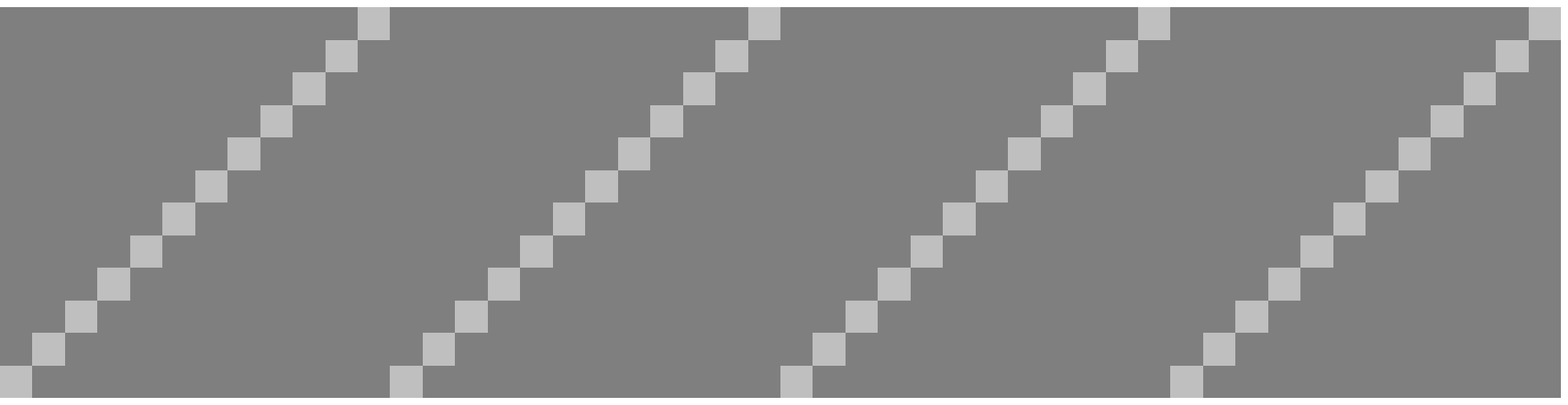}
      \end{minipage}
    \end{minipage}
    \vskip-2mm
    \caption{Constraints and null space matrices for PD with only known
      differences in phase. $K=4$ and $M=12$. White and black correspond to
      $\pm1$, respectively. See \eqsref{eq:PD-LEC} and
      (\ref{eq:PD-known-matrices}).}
    \label{fig:PD-known-matrices}
  \end{minipage}
  \hfill
  \begin{minipage}[b]{.47\textwidth}
    \centering
    \begin{minipage}[t]{.7\textwidth}
      \centering
      \begin{minipage}[t]{.6\textwidth}
        \vss
        \includegraphics[angle=-90,bb=23 35 505 758,width=\textwidth]
        {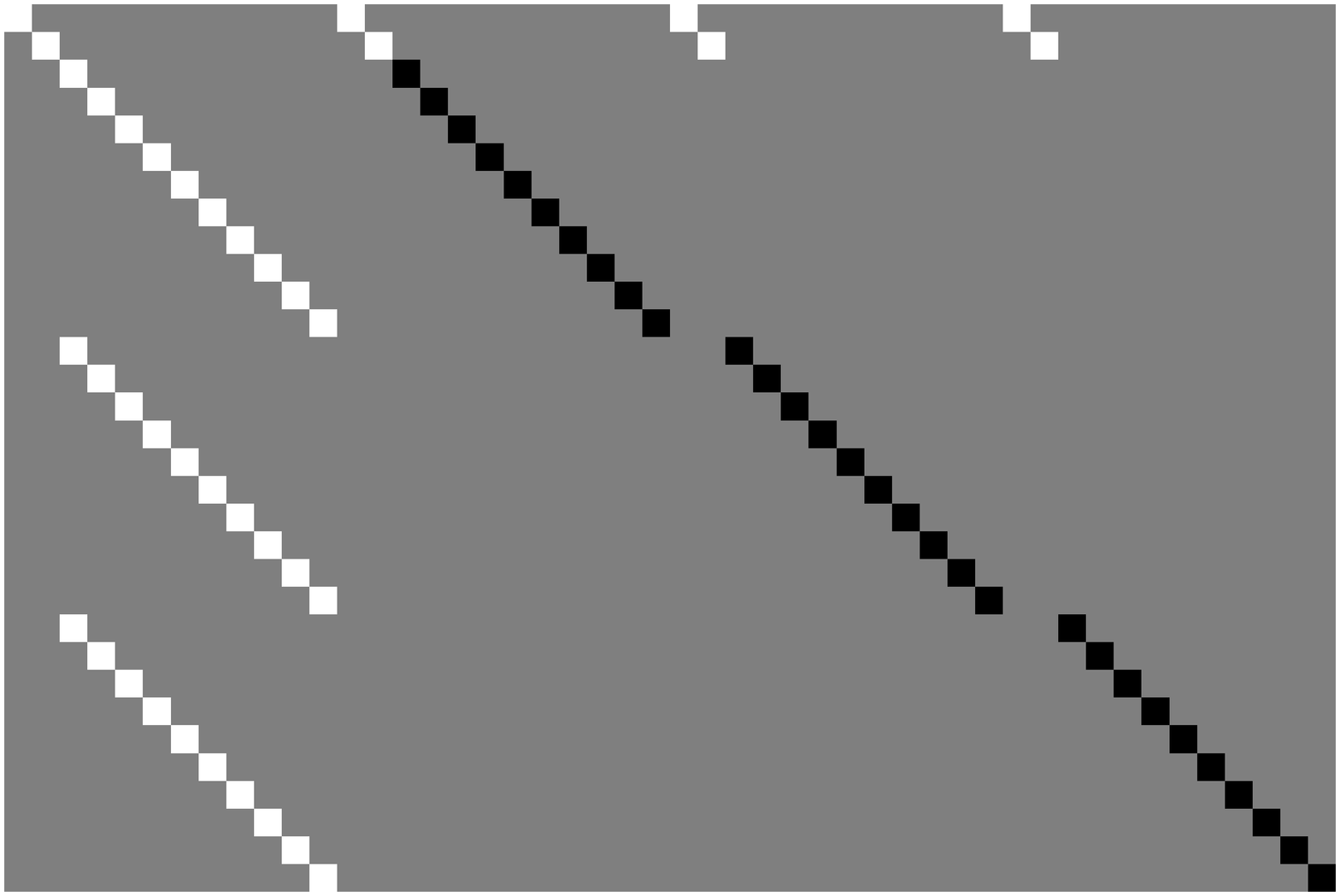}
        \vskip.15\abovecaptionskip
        \hbox{\footnotesize\textbf{(a)} $\C^{\rm PD}$}
        \vskip\belowcaptionskip
        \vfill
        \vskip0.6\abovecaptionskip
        \hfill\hbox{\footnotesize\textbf{(b)} ${\K}_2$\ }
        \vskip\belowcaptionskip
      \end{minipage} 
      \
      \begin{minipage}[t]{.2\textwidth}
        \vss
        \includegraphics[angle=-90,bb=0 35 505 205,width=\textwidth]
        {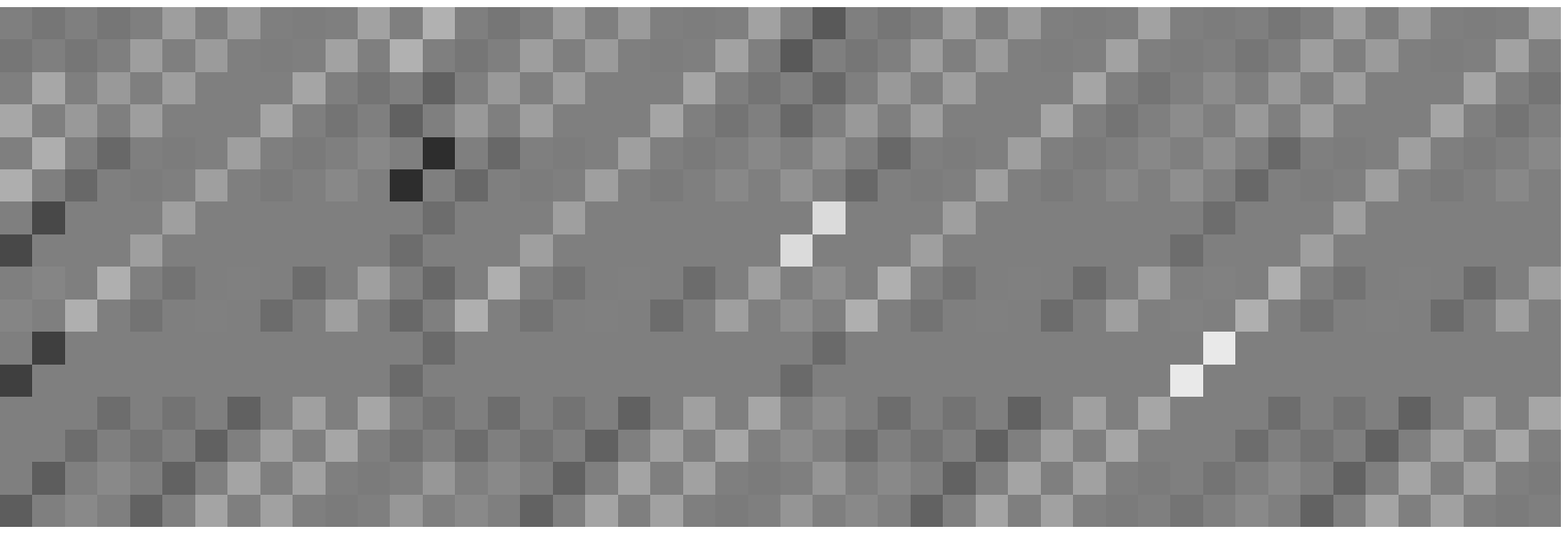}
      \end{minipage}
    \end{minipage}
    \vskip-2mm
    \caption{Constraints and null space matrices for PD with two unknown
      phase differences, e.g. the tilts. $K=4$ and $M=12$. See
      \eqref{eq:PD-unknown-LEC}. Compare
      \figref{fig:PD-known-matrices}.}
    \label{fig:PD-unknown-matrices}
  \end{minipage}
  \hfill
\end{figure}

For a PD data set ($T=1$) we know that the phases are equal, except
for diversity and registration.  The diversity is in $\theta_k$ and
does not enter in the expansion in basis functions.  To reduce the
$KM$ wavefront parameters to $M$ unknowns, we need a constraints
matrix consisting of $(K-1)M$ constraints, which can be written as
\begin{equation}
  \label{eq:PD-LEC}
  \C^{\rm PD'}:\quad\alpha_{1m}-\alpha_{km}=0; \qquad 
  k>1,\quad
  \forall m,
\end{equation}
where we momentarily disregard the registration of the imaging
channels. The rows can be in any order but it seems natural to let $m$
vary faster than $k$.  Using \eqref{eq:QR}, we get a null space
matrix.  These matrices can be written in block matrix form as
\begin{equation}
  \label{eq:PD-known-matrices}
  \C^{\rm PD'} = 
  \begin{bmatrix}
     \I_M  &-\I_M& \\
     \vdots  &   & \ddots& \\
     \I_M  &   &   & -\I_M \\
  \end{bmatrix}
  ;\qquad
  \Q_2 = \frac1{\sqrt{K}}
  \begin{bmatrix} \I_M \\\vdots\\\I_M\end{bmatrix},  
\end{equation}
where $\I_M$ is the $M\times M$ identity matrix.  $\C^{\rm PD'}$ is
a $(K-1)M \times KM$-matrix and the corresponding null space matrix is
$KM\times M$, see example in \figref{fig:PD-known-matrices}. Note that
this particular null space matrix (for every constraints matrix, there
are infinitely many) is sparse, so matrix multiplications with $\Q_2$
can be fast, even for large problems.

$\Q_2$ is easily interpreted by looking at \eqref{eq:subst}. The $K$
individual $\a_j$ are identical copies of $\B/\sqrt{K}$. Identical
$\a_j$ are exactly what we expect, given the constraints. Comparing
\eqsref{eq:L-minimization-unconstrained} and
(\ref{eq:L-minimization-constrained-PD}), it is also easy to see that
$\K_2^{\rm T}\cdot\b^{\rm MFBD} = \b^{\rm PD}$ (except for an
inconsequential constant factor in the wavefront regularization term),
so the same solution minimizes the problem in either formulation. For
each specific $K$ it is also easy to confirm that
$\K_2^{\rm T}\A^{\rm MFBD}\K_2=\A^{\rm PD}$. So the entire original
PD problem from \secref{sec:trad-phase-diversity} is
retained.

\subsection{Phase Diversity with Partly Unknown Differences in Phase}
\label{sec:PD-unknown}

For modes with unknown differences between the channels, we simply do
not add any constraint, thus allowing the algorithm to optimize the
corresponding coefficients independently in the two channels based on
the available data.
This can be used e.g. for the focus mode, if the diversity it is not
known well, or for other low order modes that may differ between
diversity channels.

The most common unknown differences, however, are for the tilt modes,
corresponding to image registration. Unless we have information from
other data sets, we generally don't know the tilt differences. So we
want to exclude the tilt modes from Eq.~(\ref{eq:PD-LEC}).  However,
we may wish to prevent a common tilt to grow without bounds by
requiring that they add up to zero,
\begin{equation}
  \C^{\rm PD}:\quad
  \begin{cases}
    \sum_k \alpha_{km} = 0;& \qquad m\in\{{\rm tilt\ modes}\} \\
    \alpha_{1m}-\alpha_{km}=0;& \qquad  k>1,\quad m \not\in\{{\rm tilt\ modes}\}.
    \label{eq:PD-unknown-LEC}
  \end{cases}
\end{equation}
This will still allow changes in the individual tilt coefficients. A
sample $((K-1)(M-2)+2) \times KM$ constraints matrix and a
corresponding null space matrix are shown in
\figref{fig:PD-unknown-matrices}. The modes with unknown differences
(tilts) are numbered as $m\in\{1,2\}$.  The null space matrix
generated by my QR factorization code is not very regular but it does
work, however it is not as easily interpreted as the one in the
previous subsection. Also, it is not as sparse. See discussion in
\secref{sec:discussion}.

The number of unknowns is the $M-2$ non-tilt wavefront parameters for
the common phase plus the registration of $K$ channels, which amounts
to $2(K-1)$ tilt parameters.  This corresponds to the total number of
MFBD wavefront parameters, $KM$, minus the number of constraints,
$(K-1)(M-2)+2$.

\subsection{Phase Diverse Speckle}
\label{sec:JPDS}

\begin{figure}[tbp]
  \centering
  \hfill
  \begin{minipage}[b]{.47\textwidth}
    \centering
    \begin{minipage}[t]{.9\textwidth}
      \centering
      \begin{minipage}[t]{.6\textwidth}
        \vss
        \includegraphics[angle=-90,bb=142 34 505 758,width=\textwidth]
        {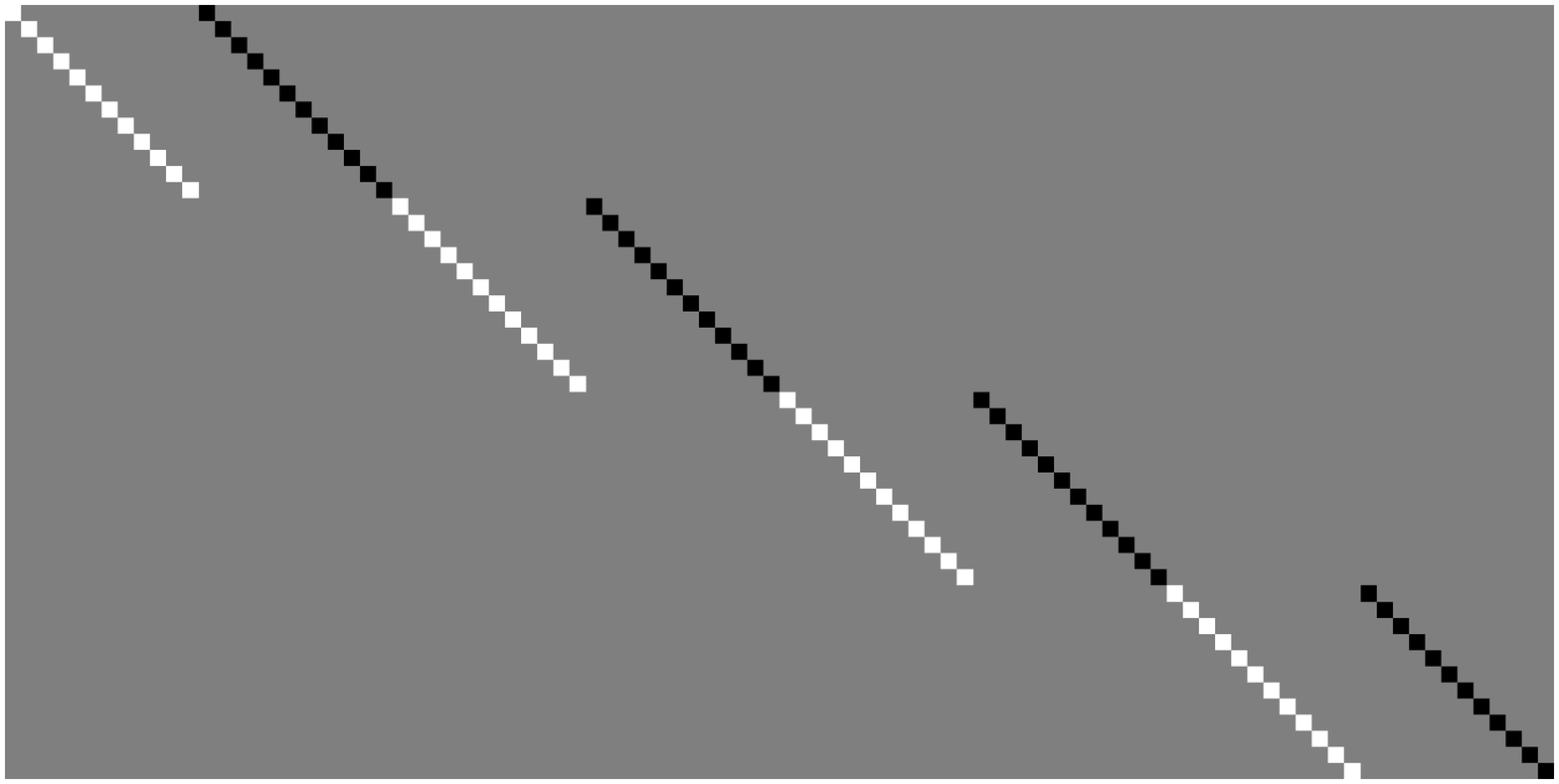}
        \vskip.15\abovecaptionskip
        \hbox{\footnotesize\textbf{(a)} $\C^{\rm PDS'}$}
        \vskip\belowcaptionskip
        \vfill
        \vskip3\abovecaptionskip
        \hfill\hbox{\footnotesize\textbf{(b)} ${\K}_2$\ }
        \vskip\belowcaptionskip
      \end{minipage}
      \
      \begin{minipage}[t]{.3\textwidth}
        \vss
        \includegraphics[angle=-90,bb=0 35 505 290,width=\textwidth]
        {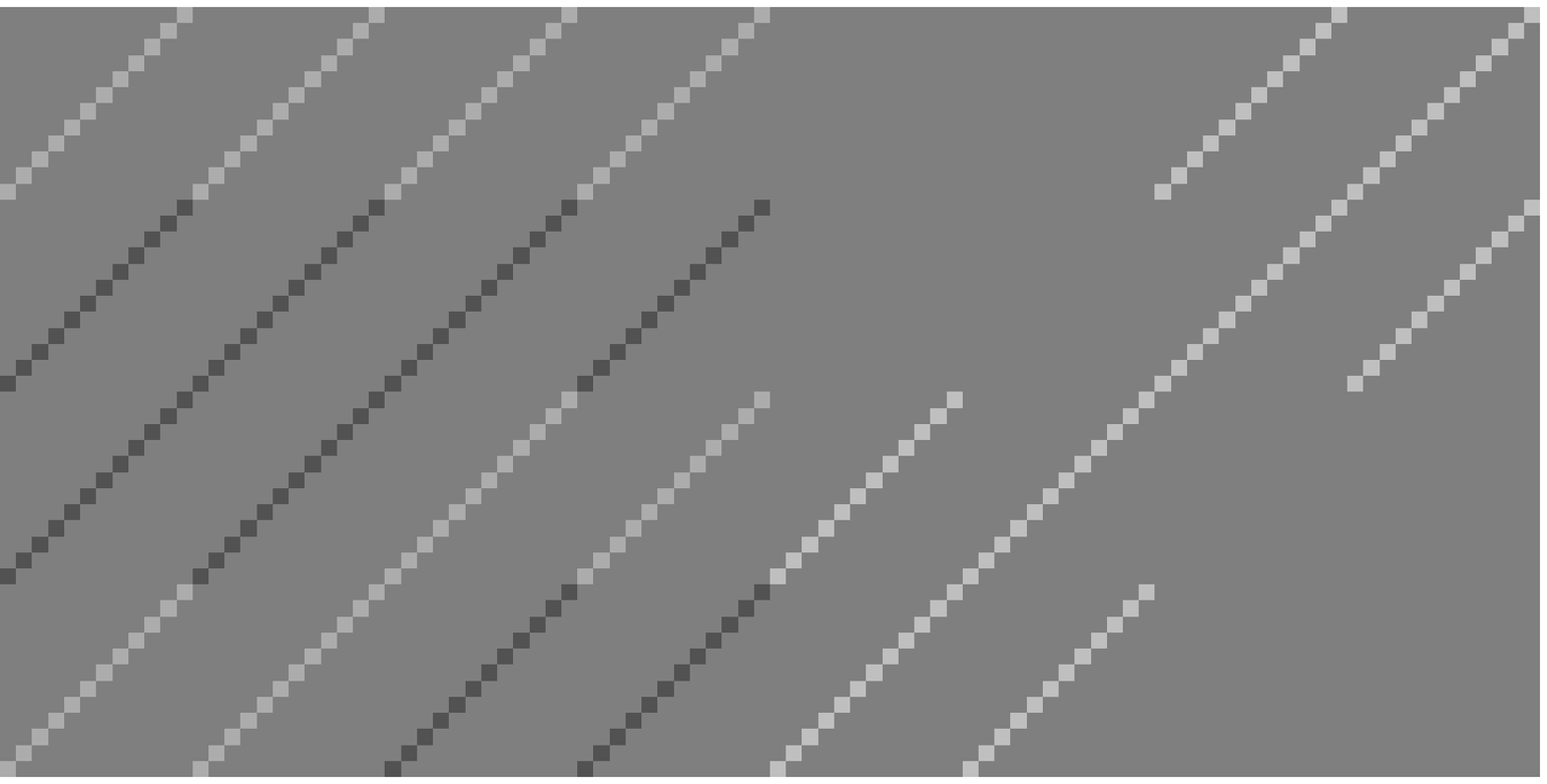}
      \end{minipage}
    \end{minipage}
    \vskip-2mm
    \caption{Constraints and null space matrices for PDS with only known
      differences in phase. $J=4$, $K=2$ and $M=12$. See
      \eqref{eq:PDS-unknown-LEC}, ignore special treatment of tilt
      modes.}
    \label{fig:PDS-known-matrices}
  \end{minipage}
  \hfill
  \begin{minipage}[b]{.47\textwidth}
    \centering
    \begin{minipage}[t]{.9\textwidth}
      \centering
      \begin{minipage}[t]{.6\textwidth}
        \vss
        \includegraphics[angle=-90,bb=142 34 505 758,width=\textwidth]
        {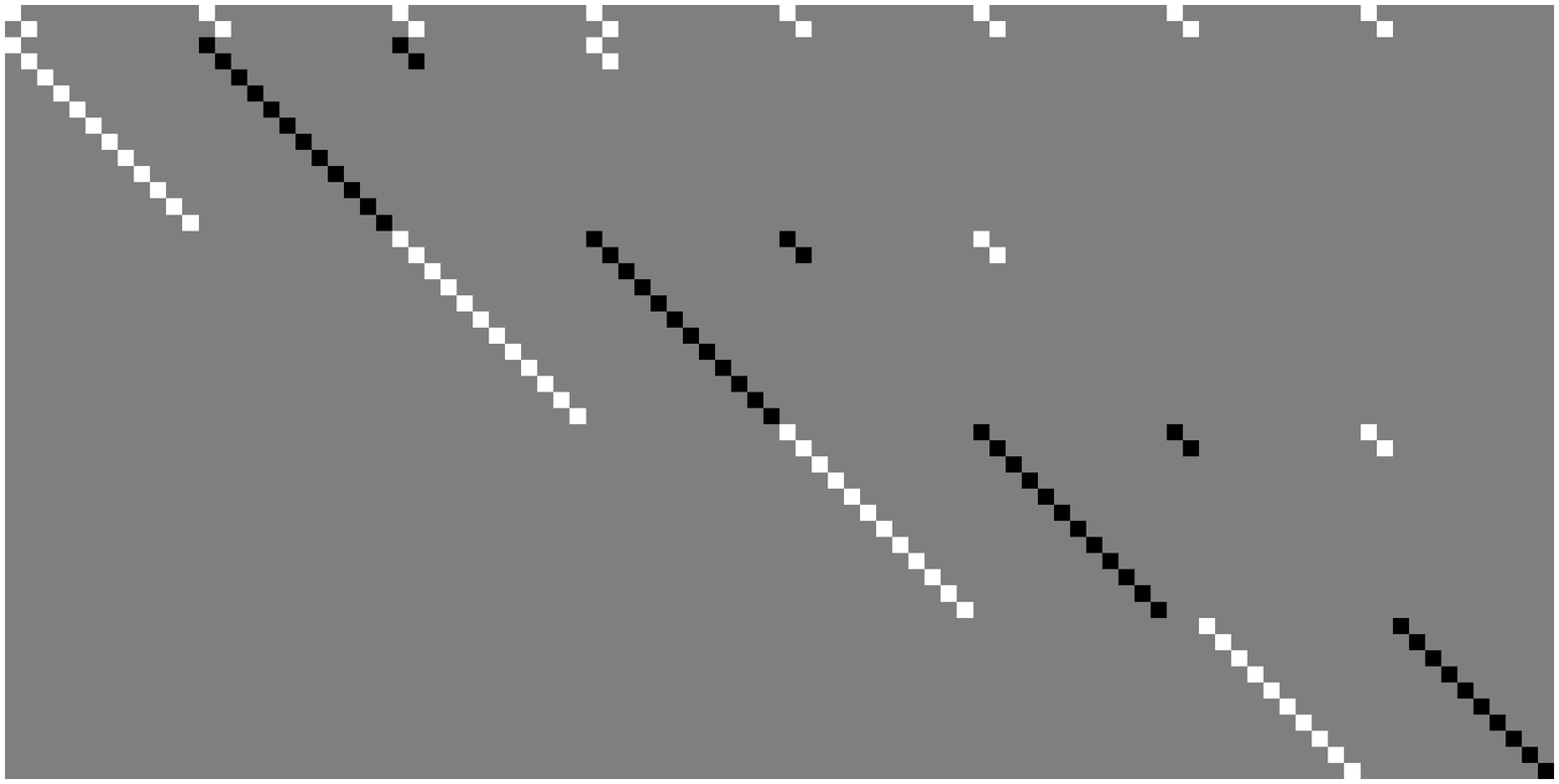}
        \vskip.15\abovecaptionskip
        \hbox{\footnotesize\textbf{(a)} $\C^{\rm PDS}$}
        \vskip\belowcaptionskip
        \vfill
        \vskip3\abovecaptionskip
        \hfill\hbox{\footnotesize\textbf{(b)} ${\K}_2$\ }
        \vskip\belowcaptionskip
      \end{minipage} 
      \
      \begin{minipage}[t]{.3\textwidth}
        \vss
        \includegraphics[angle=-90,bb=0 35 505 290,width=\textwidth]
        {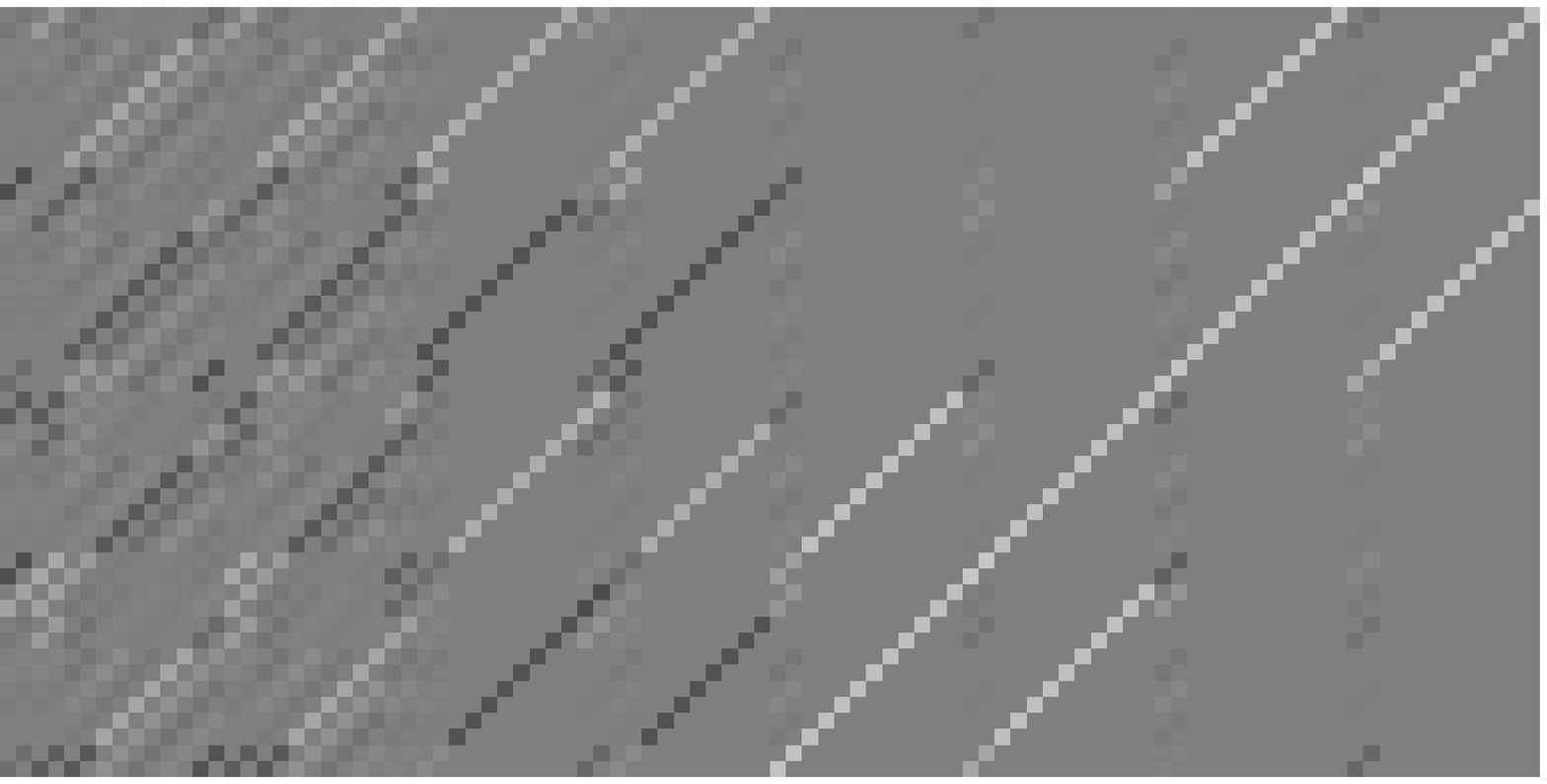}
      \end{minipage}
    \end{minipage}
    \vskip-2mm
    \caption{Constraints and null space matrices for PDS with two unknown
      phase differences, e.g. the tilts. $J=4$, $K=2$ and $M=12$. See
      \eqref{eq:PDS-unknown-LEC}.}
    \label{fig:PDS-unknown-matrices}
  \end{minipage}
\end{figure}

Again starting with the case of completely known phase differences,
the number of wavefront parameters is $KTM$, but there are only $T$
different atmospheric realizations. The real number of unknowns is
therefore $TM$ and we need $(K-1)TM$ constraints.  This is $T$ times
the number of constraints in the PD problem.  For each
$t\in\{1,\ldots,T\}$ we simply add one set of constraints of the
second type in \eqref{eq:PD-unknown-LEC}, see sample matrices in
\figref{fig:PDS-known-matrices}.

In practice, we must allow for unknown registration between PD
channels as well as between consecutive PD sets. Normally we know that
the difference in tilt between the PD channels, $k$, is constant over
time, $t$. Finally, we require that the registration parameters sum up
to zero independently in the two directions.
\begin{equation}
  \label{eq:PDS-unknown-LEC}
  \C^{\rm PDS}:\quad
  \begin{cases}
    \sum_{t,k} \alpha_{tkm} = 0; & \qquad m\in\{{\rm tilt\ modes}\}\\
    \alpha_{tkm} - \alpha_{t(k-1)m} = \alpha_{(t-1)km} - \alpha_{(t-1)(k-1)m};
    & \qquad  t>1, \quad k>1,\quad m\in\{{\rm tilt\ modes}\}\\
    \alpha_{t1m}-\alpha_{tkm}=0; & \qquad 
    \forall t,\quad k>1,\quad
    m \not\in\{{\rm tilt\ modes}\}.
  \end{cases}
\end{equation}
Note that this problem is just as constrained as the problem with
completely known phase differences, thus the null space matrices have
the same dimensions.  See sample matrices in
\figref{fig:PDS-unknown-matrices}. The special tilt constraints result
in a less orderly and sparse null space matrix, although not as bad as
in the PD case.

When working with seeing degraded data, it is likely that all PD sets
are not of the same quality. One strength with PDS is that bad data
can be helped by being processed jointly with the worse data. However,
intuition suggests that there are cases where the bad data disturbs
the inversion of the good data.  It can then be useful to calculate
individual metrics for each PD set, use \eqref{eq:L-mfbd} but skip the
$\gamma_{\rm wf}$ term and limit the sum over $j$ to one $t$.
\begin{equation}
  \label{eq:L-mfbd-individual}
  L_t = \sum_u\biggl[
  \sum_k^K 
  \abs{D_{tk}}^2 -
  \frac{\abs{\sum_k^K D^*_{tk} S_{tk}}^2}{Q_t}
  \biggr],
\end{equation}
where the sum in $Q_t$ is limited correspondingly.  This gives good
diagnostics for determining which PD sets resulted in good wavefront
estimates.

\subsection{Shack--Hartmann WFS}
\label{sec:shack-hartmann-wfs}

A Shack--Hartmann (SH) wavefront sensor consists of a microlens array
that form a number of sub-images, each sampling different parts of the
pupil. SH wavefront sensing works with point sources as well as with
extended objects. An example of the latter is that it has been used
with solar telescopes\cite{rimmele97high,scharmer00workstation}.  A
fundamental limitation for such applications is that the number of
microlenses is restricted by the requirement that each sub-image must
resolve solar fine structure.

Usually, the mean gradient of the phase is estimated by measuring the
relative or absolute image motion in the subimages. These local tilts
are then combined into a phase over the whole pupil.  This means that
in conventional analysis of SH wavefront data, information about
wavefront curvature over the microlenses, manifested as differential
blurring in the subimages, is discarded. It is only by more detailed
modeling that all relevant information on the wavefronts can be
extracted, which in principle should be made down to the noise level
of the data.

With different $A_j$ for different imaging channels, $j$,
\eqsref{eq:pupil}--(\ref{eq:phi}) accommodate SH data. The phase over
the full pupil is constrained to be the same for all channels, while
the different $A_j$ correspond to the different parts of the pupil
that are sampled by the different microlenses. The constraints matrix
is the same as for PD with $J$ channels.

Calibration of the relative positions of the SH sub-images is required
for conventional processing. If such data are available, they can be
entered as tilts in the different $\theta_j$. However, it should be
possible to estimate these tilts from the data if $\theta_j$ is set to
zero and local tilts are included in the parameterization of the
wavefront. On the other hand, no global tilts are necessary.

It has been demonstrated in simulation with two microlenses that this
image formation model allows higher modes to be estimated than with
conventional local tilt approximations, because local aberrations
within each subimage is taken into account\cite{lofdahl98fast}. SH
data have also been considered for regular PD analysis.  The microlens
array acts as a single phase screen and the sub-images are regarded as
a single phase diverse image, to be processed together with a
conventional high-resolution
image\cite{paxman93fine-resolution-SPIE,schulz93estimation}. Both
approaches are sensitive to getting the geometry of the microlens
array right (phase screen in the latter case, positions of the $A_j$
in the MFBD--LEC case). However, in the MFBD--LEC approach, the
registration of the subimages does not have to be known; it can be
estimated.

You can still use a conventional image as another image channel in the
MFBD--LEC formulation. If local tilt calibration data is available,
this should be straight-forward. One would want to avoid estimating
local microlens tilts for the high-resolution channel. This can easily
be done by adding LECs that set them to zero. See also
\secref{sec:relax-M}.

\section{Extending the presented formulation}

The formulation presented so far was chosen so that it would be fairly
straight-forward, while still accommodating PDS and SH WFS.  In order
to demonstrate the versatility of the LEC approach, we now discuss
briefly a number of relaxations of some of the requirements, and the
types of data sets that can then be treated.

\subsection{Different Phase Parameterization}
\label{sec:relax-M}

It is trivial to relax the requirements that $M$ is the same for every
data set. The single index of the $\a$ and $\b$ arrays can then be
written as $n=m + \sum_{j'=1}^{j-1}M_{j'}$ rather than $n=JM$.  

One application for this is the SH processing of
\secref{sec:shack-hartmann-wfs}. If the SH sub-images are to be
processed together with a conventional high-resolution image, one may
want to exclude the local sub-pupil tilts from the parameterization of
the phase seen by the conventional camera. See also the following
subsections.

\subsection{Different Objects}
\label{sec:relax-object}

We can also relax the requirement that all images have the same
object. This requires involving another index, say $s\in\{1,\dots,S\}$
(for ``scene'' or ``set''), and changing the formulae for the metric
and its derivatives to allow for separate objects $F_s$. The metric
can then be written as $L=\sum_s L_s$, where $L_s$ is from
\eqref{eq:L-mfbd} but summing over $j$ only that involves object
$F_s$.

One application for this is high-resolution solar magnetograms, see
Ref.\@~\citenum{lundstedt91magnetograph,lites91xx}. Magnetograms can
be made by calculating the difference between opposite polarization
components of light in a Zeeman sensitive spectral line collected
through a birefringent filter. Because the magnetic signal is
essentially the difference between two very similar images, it is
important to minimize artifacts from a number of error sources, among
them registration, seeing, in particular differential seeing, and
instrumental effects.

The differential seeing problem can be handled by making each image
the sum of many short exposures, where bad frames are discarded and
the images can be well registered before adding. The short exposures
can be made simultaneously in the two polarization channels by using a
polarizing beamsplitter. To reduce also the influence of different
aberrations in the paths after the beamsplitter, the polarization
states can be rapidly switched between the detectors, so that each
polarization state is recorded with a aberrations from both paths.

In order to reduce the effects of seeing degradation of the image
resolution, the two sequences of short exposures could also be MFBD
restored for seeing effects. Better would be to do this using
constraints that come from the fact that the aberration differences
between the two channels is constant, with proper respect paid to the
switching.  Since registration parameters are estimated along with the
phase curvature terms, this could solve the seeing problem while at
the same time registering the image channels to subpixel accuracy.
This should be better than registration with cross-correlation
techniques, since non-symmetric instantaneous PSFs would tend to
corrupt such alignment.

The fact that the differences between the two polarization signals are
so subtle might inspire another approach. If the object is considered
to be the same for wavefront sensing purposes, then we can treat the
data as a PDS data set with zero diversity terms $\theta_j$. The
object is the same and so are the phases, except for the registration
terms.

\subsection{Different wavelengths}
\label{sec:relax-wavelength}

If PDS data of the same object is collected in several different
wavelengths, and the cameras are synchronized, it has been shown that
wavefronts estimated from one PDS set can be used for restoration of
the object seen in the data collected in the other
wavelengths\cite{lofdahl01two}. However, this requires information
about the wavefront differences as seen by the different cameras. The
differences that come from the different wavelengths can easily be
calculated and other differences can be estimated from a selected
subset of a larger data set and then applied to all the data under the
assumption that they do not change.

This assumption could also be used to advantage in a joint treatment
of the simultaneous PDS (SPDS) sets.  The different-object
generalization from \secref{sec:relax-object} is used together with
PDS LECs and LECs that express this assumption,
\begin{equation}
  \label{eq:SPDS-unknown-LEC}
  \C^{\rm SPDS}:\quad
  \begin{cases}
    \sum_{s,t,k} \alpha_{stkm} = 0; & \qquad \forall s,\quad
    m\in\{{\rm tilt\ modes}\}\\
    \alpha_{stkm} - \alpha_{st(k-1)m} = \alpha_{s(t-1)km} - \alpha_{s(t-1)(k-1)m};
    & \qquad \forall s,\quad  t>1, \quad k>1,\quad m\in\{{\rm tilt\ modes}\}\\
    \alpha_{st1m}-\alpha_{stkm}=0; & \qquad 
    \forall s,\quad    \forall t,\quad k>1,\quad
    m \not\in\{{\rm tilt\ modes}\}\\
  \lambda_{1}(\alpha_{111m}-\alpha_{1t1m})= 
  \lambda_{s}(\alpha_{s11m}-\alpha_{st1m});
  & \qquad s>1,\quad t>1,\quad m \not\in\{{\rm tilt\ modes}\}
  \end{cases}
\end{equation}
where $\lambda_{s}$ is the wavelength (in arbitrary units)
used for set $s$.  For the assumption of wavelength independent
aberrations to hold, the optical system should be approximately
achromatic. In practice, this means the differences between the
wavelengths should not be too large.

Again, this can also be relevant for SH, if a high-resolution image is
processed together with the SH data -- and the wavelength is
not the same in the SH sensor as in the imaging camera.

\subsection{Different Numbers of Diversity Channels}
\label{sec:relax-K}

Finally, we can also treat data sets with different numbers of phase
diversity channels in different wavelengths. This requires a $K$ that
varies with $s$, $J=\sum_{s} K_s$. This is perhaps most useful because
it permits setting $K_s=1$ for some $s$. This corresponds to PDS data
set in one wavelength and simultaneous MFBD data in another wavelength
as in Refs.\@~\citenum{paxman99phase-diversity,tritschler02sunspot}.

We should be able to run several such sets jointly and constrain
the aberration differences between different $s$ to be the same
for all $t$, without requiring the object to be the same.

Again, the advantage of the joint treatment is of course that
wavefront differences between different imaging cameras do not have to
be calibrated but are estimated together with the aberrations that are
in common to the different cameras.

\section{Discussion}
\label{sec:discussion}

We have presented a formulation of the MFBD problem that accomodates
the varying data collection schemes involved in PD and PDS as well as
SH WFS and a number of combinations of data in different wavelengths
or polarizations and with or without diversity in the phase or pupil
geometry.  In doing this, we have not exhausted the types of data sets
that can be treated with the method presented in this paper.  As long
as there are known relations between the imaging channels, that can be
formulated in terms of constraints on the wavefront coefficients, the
data could be handled and the constraints can be used to benefit
wavefront sensing and image restoration.

We have written a code that implements the method presented in this
paper. It has been tested for PD, PDS and MFBD, but the other
suggested processing strategies have not been tried yet. Trying some
of the suggested strategies will be the subject of future papers.

The code is new and as we apply it to different problems we anticipate
significant improvements in the implementation of the method as well
as in the method itself. One particular area of improvement is the
calculation of the null space matrix. It consists of basis vectors for
the null space and for any constraints matrix, there are infinitely
many null space matrices, corresponding to rotations of the coordinate
system in the null space.  A method for making a null space matrix
that is as sparse and regular as possible would be very useful.
Matrix multiplications involving sparse matrices can be performed much
faster than full matrices when the size of the problem is large. Also,
a null space matrix such as the one in \figref{fig:PD-known-matrices}
is so much more instructive and easy to interpret than the one in
\figref{fig:PD-unknown-matrices}.


\section*{Acknowledgments}       

I am grateful to Gerd and Henrik Eriksson at the Royal Institute of
Technology in Stockholm for helpful comments and advice on many of the
mathematical and numerical concepts used in this paper, including in
particular constrained linear equations systems and null spaces.  I am
grateful to Curtis Vogel and Luc Gilles of Montana State University
for correspondence on Hessians and optimization methods.

This research was supported in part by Independent Research and
Development funds at Lockheed Martin Space Systems, Advanced
Technology Center and by the MDI project at Stanford and Lockheed
Martin Solar and Astrophysics Laboratory, NASA grant NAG5-3077.


\bibliography{bib-strings,svst,mats,mats.lofdahl}

\begin{thebibliography}{10}

\bibitem{schulz93multi-frame}
T.~J. Schulz, ``Multi-frame blind deconvolution of astronomical images,'' {\em
  Journal of the Optical Society of America A} {\bf 10}, pp.~1064--1073, 1993.

\bibitem{schulz97multiframe}
T.~J. Schulz, B.~E. Stribling, and J.~J. Miller, ``Multiframe blind
  deconvolution with real data: Imagery of the {H}ubble {S}pace {T}elescope,''
  {\em Optics Express} {\bf 1}(11), pp.~355--362, 1997.

\bibitem{kampen98multi-frame}
W.~C. Van~Kampen and R.~G. Paxman, ``Multi-frame blind deconvolution of
  infinite-extent objects,'' in {\em Propagation and Imaging through the
  Atmosphere II},  L.~R. Bissonnette, ed., {\em Proc. SPIE} {\bf 3433},
  pp.~296--307, 1998.

\bibitem{tyler98comparison}
D.~W. Tyler, S.~D. Ford, B.~R. Hunt, R.~G. Paxman, M.~C. Roggeman, J.~C.
  Roundtree, T.~J. Schulz, K.~J. Schulze, J.~H. Seldin, G.~G. Sheppard, B.~E.
  Stribling, W.~C. Van~Kampen, and B.~M. Welsh, ``Comparison of image
  reconstruction algorithms using adaptive optics instrumentation,'' in {\em
  Adaptive Optical System Technologies},  D.~Bonaccini and R.~K. Tyson, eds.,
  {\em Proc. SPIE} {\bf 3353}, pp.~160--171, 1998.

\bibitem{paxman92phase}
R.~G. Paxman, T.~J. Schulz, and J.~R. Fienup, ``Phase-diverse speckle
  interferometry,'' in {\em Signal Recovery and Synthesis {IV}},  {\em
  Technical Digest Series} {\bf 11}, pp.~5--7, Optical Society of America,
  1992.

\bibitem{lofdahl94wavefront}
M.~G. L{\"o}fdahl and G.~B. Scharmer, ``Wavefront sensing and image restoration
  from focused and defocused solar images,'' {\em Astronomy \& Astrophysics
  Supplement Series} {\bf 107}, pp.~243--264, 1994.

\bibitem{paxman96evaluation}
R.~G. Paxman, J.~H. Seldin, M.~G. L{\"o}fdahl, G.~B. Scharmer, and C.~U.
  Keller, ``Evaluation of phase-diversity techniques for solar-image
  restoration,'' {\em Astrophysical Journal} {\bf 466}, pp.~1087--1099, 1996.

\bibitem{gonsalves82phase}
R.~A. Gonsalves, ``Phase retreival and diversity in adaptive optics,'' {\em
  Optical Engineering} {\bf 21}(5), pp.~829--832, 1982.

\bibitem{paxman92joint}
R.~G. Paxman, T.~J. Schulz, and J.~R. Fienup, ``Joint estimation of object and
  aberrations by using phase diversity,'' {\em Journal of the Optical Society
  of America A} {\bf 9}(7), pp.~1072--1085, 1992.

\bibitem{lofdahl00calibration}
M.~G. L{\"o}fdahl, G.~B. Scharmer, and W.~Wei, ``Calibration of a deformable
  mirror and {S}trehl ratio measurements by use of phase diversity,'' {\em
  Applied Optics} {\bf 39}(1), pp.~94--103, 2000.

\bibitem{vogel98fast}
C.~R. Vogel, T.~F. Chan, and R.~J. Plemmons, ``Fast algorithms for phase
  diversity-based blind deconvolution,'' in {\em Adaptive Optical System
  Technologies},  D.~Bonaccini and R.~K. Tyson, eds., {\em Proc. SPIE} {\bf
  3353}, 1998.

\bibitem{engl96regularization}
H.~W. Engl, M.~Hanke, and A.~Neubauer, {\em Regularization of Inverse
  Problems}, vol.~375 of {\em Mathematics and Its Applications}, Kluwer
  Academic Publishers, Dordrecht, Netherlands, 1996.

\bibitem{roddier90atmospheric}
N.~Roddier, ``Atmospheric wavefront simulation using {Z}ernike polynomials,''
  {\em Optical Engineering} {\bf 29}(10), pp.~1174--1180, 1990.

\bibitem{kahaner89numerical}
D.~Kahaner, C.~Moler, and S.~Nash, {\em Numerical Methods and Software},
  Prentice Hall, 1989.

\bibitem{rimmele97high}
T.~Rimmele, J.~M. Beckers, R.~B. Dunn, R.~R. Radick, and M.~Roeser, ``High
  resolution solar observations from the ground,'' in {\em The High Resolution
  Solar Atmospheric Dynamics Workshop},  {\em PAPSP conference series}, 1997.

\bibitem{scharmer00workstation}
G.~B. Scharmer, M.~Shand, M.~G. L{\"o}fdahl, P.~M. Dettori, and W.~Wei, ``A
  workstation based solar/stellar adaptive optics system,'' in {\em Adaptive
  Optical Systems Technologies},  P.~L. Wizinowich, ed., {\em Proc. SPIE} {\bf
  4007}, pp.~239--250, 2000.

\bibitem{lofdahl98fast}
M.~G. L{\"o}fdahl, A.~L. Duncan, and G.~B. Scharmer, ``Fast phase diversity
  wavefront sensing for mirror control,'' in {\em Adaptive Optical System
  Technologies},  D.~Bonaccini and R.~K. Tyson, eds., {\em Proc. SPIE} {\bf
  3353}, pp.~952--963, 1998.

\bibitem{paxman93fine-resolution-SPIE}
R.~G. Paxman and J.~H. Seldin, ``Fine-resolution astronomical imaging with
  phase-diverse speckle,'' in {\em Digital Recovery and Synthesis {II}},  P.~S.
  Idell, ed., {\em Proc. SPIE} {\bf 2029}, pp.~287--298, 1993.

\bibitem{schulz93estimation}
T.~J. Schulz, ``Estimation-theoretic approach to the deconvolution of
  atmospherically degraded images with wavefront sensor measurements,'' in {\em
  Digital Recovery and Synthesis {II}},  P.~S. Idell, ed., {\em Proc. SPIE}
  {\bf 2029-31}, 1993.

\bibitem{lundstedt91magnetograph}
H.~Lundstedt, A.~Johannesson, G.~Scharmer, J.~O. Stenflo, U.~Kusoffsky, and
  B.~Larsson, ``Magnetograph observations with the {S}wedish solar telescope on
  {La~Palma},'' {\em Solar Physics} {\bf 132}, pp.~233--245, 1991.

\bibitem{lites91xx}
B.~W. Lites in {\em Solar Polarimetry},  L.~J. November, ed., {\em Proc. 11th
  Sacramento Peak Summer Workshop}, p.~173, 1991.

\bibitem{lofdahl01two}
M.~G. L{\"o}fdahl, T.~E. Berger, and J.~H. Seldin, ``Two dual-wavelength
  sequences of high-resolution solar photospheric images captured over several
  hours and restored by use of phase diversity,'' {\em Astronomy \&
  Astrophysics} {\bf 377}, pp.~1128--1135, 2001.

\bibitem{paxman99phase-diversity}
R.~G. Paxman and J.~H. Seldin, ``Phase-diversity data sets and processing
  strategies,'' in {\em High Resolution Solar Physics: Theory, Observations and
  Techniques},  T.~Rimmele, R.~R. Radick, and K.~S. Balasubramaniam, eds., {\em
  Proc. 19th Sacramento Peak Summer Workshop, ASP Conf. Series vol. 183},
  pp.~311--319, 1999.

\bibitem{tritschler02sunspot}
A.~Tritschler and W.~Schmidt, ``Sunspot photometry with phase diversity. {I}.
  {M}ethods and global sunspot parameters,'' {\em Astronomy \& Astrophysics}
  {\bf 382}, pp.~1093--1105, 2002.

\end{thebibliography}

\end{document}